\documentclass[10pt]{article}

\usepackage{graphicx}

\newcommand{\bs}{\mathbf}

\newtheorem{thm}{Theorem}[section]

\newtheorem{prop}[thm]{Proposition}

\newenvironment{rem}{\refstepcounter{thm} \bigskip \par \noindent
    {\bf Remark \thethm}\ }{\bigskip \par}
\newenvironment{dfn}{\refstepcounter{thm} \bigskip \par \noindent
    {\bf Definition \thethm}\ }{\bigskip \par}
\newenvironment{exa}{\refstepcounter{thm} \bigskip \par \noindent
        {\bf Example \thethm}\ }{\bigskip \par}
\newenvironment{pf}{\medskip \par \noindent {\it Proof.}\ }{\hfill
        \P \bigskip \par}

\begin{document} 
\baselineskip=15pt

\title{Optimal exact tests for composite alternative hypotheses on cross tabulated data}

\author{Daniel Yekutieli}

\maketitle

\begin{abstract}

We present methodology for constructing exact significance tests for cross tabulated data 
for ``difficult'' composite alternative hypotheses that have no natural test statistic.
We construct a test for discovering Simpson's Paradox and a general test for discovering positive dependence between two ordinal variables.
Our tests are Bayesian extensions of the likelihood ratio test,
they are optimal with respect to the prior distribution,
and are also closely related to Bayes factors and Bayesian FDR controlling testing procedures.
\end{abstract}

\section{Introduction}

We present  Bayesian extensions of the likelihood ratio test
that are optimal with respect to the prior distribution
for testing composite alternative hypotheses that have no natural test statistic.
As a motivating example, we present exact tests for discovering Simpson's paradox.

\begin{exa} \label{exa11} 
Table 1 displays data from a study on Death penalty in Florida (Agresti 2002,  Table 2.13).
The $326$ subjects classified in Table 1 were the defendants in indictments involving cases with multiple 
murders in Florida. 
The goal of the analysis is to determine whether the probability of receiving death sentence depends on the defendant's race.

 The variables are $X$ -- Race of Victim (``White'', ``Black''),
$Y$ --  Race of Defendant  (``White'', ``Black'')'), and $Z$ -- Death Penalty verdict (``Yes'', ``No"). 
$\pi_{i j k}$ is the probability 
that $X$ takes on its $i$th value and $Y$ takes on its $j$th value  and $Z$ takes on its $k$th value.
The conditional odds ratio between defendant's race and death penalty for White victims  is
$\theta_{ Y  Z | X = 1} = (\pi_{1 1 1} \cdot \pi_{1 2 2 }) /  (\pi_{1 1 2} \cdot \pi_{1 2 1})$
and for Black victims  it is
$\theta_{ Y  Z | X = 2} = (\pi_{2 1 1} \cdot \pi_{2 2 2 }) /  (\pi_{2 1 2} \cdot \pi_{2 2 1})$.
The marginal odds ratio between defendant's race and death penalty  is
$\theta_{Y  Z} = (\pi_{+ 1 1} \cdot \pi_{+ 2 2 }) /  (\pi_{+ 1 2} \cdot \pi_{+ 2 1})$,
for $\pi_{+ j k} = \pi_{1 j k} + \pi_{2 j k}$. 
Similarly, $\theta_{X  Z}$ is the marginal odds ratio between 
victim's race and death penalty and $\theta_{X Y}$ is the 
marginal odds ratio between defendant's race and death penalty.

We  used the R {\em fisher.test} function to test dependency between the pairs of variables.
Defendant race and victim race are highly dependent, $\hat{\theta}_{X Y}  = 27.1$ with  $0.95$ CI $ [12.7,  64.8 ]$;
and risk of receiving death penalty is higher for white victims than for black victims,
$\hat{\theta}_{X Z }  = 2.87$ with $0.95$ CI $[1.13,  8.73 ]$.
Thus Victim's race is a confounder:
white defendants have higher probability of receiving death penalty just because they are more likely to kill a white victim.
Indeed, we see that $\hat{\theta}_{Y Z}  = 1.18$ with $0.95$ CI $ [0.56,  2.52 ]$.
The null hypothesis we consider is that conditional on victim's race
defendant's race and death penalty are independent,
$H_0:  \;  \theta_{ Y  Z | X = 1}   = 1 , \;  \theta_{ Y  Z | X = 2}   = 1 $.
The alternative hypothesis is that the following Simpson's paradox occurs,
$H_1: \;  \theta_{ Y  Z | X = 1}  < 1,\;  \theta_{ Y  Z | X = 2}  < 1, \; 1 < \theta_{ Y  Z }$.

\medskip
To test the null hypothesis, for white victims we further condition on the observed values
 $N_{1 1 +} = 151$, $N_{1 2 +} = 63$, $N_{1 + 1} = 30$, $N_{1 + 2} = 184$,
and for Black victims we further condition  on the observed values
$N_{2 1 +} = 9$, $N_{2 2 +} = 103$, $N_{2 + 1} = 6$, $N_{2 + 2} = 106$.
Forming a conditional sample space with $217$ points that can be expressed
\[
\Omega_a = \{  ( N_{111}, N_{211}) :  \   N_{1 1 1} \in (0, 1, \cdots, 30), \ N_{2 1 1} \in (0, 1, \cdots, 6) \}.
\]
The observed data point is $(N_{111} = 19, N_{211} = 0)$.
Under $H_0$, $N_{1 1 1}$ and   $N_{2 1 1}$ are independent
and, using R notations, the probability of  each data point is
\[
\Pr_{H_0} ( N_{1 1 1} = x,  N_{2 1 1} = y  ) =  dhyper( x ; 151, 63, 30)  \cdot dhyper( y ; 9, 103, 6) \cdot 
\]
Applying the R {\em fisher.test} function to the observed 2-by-2 tables corresponding to White and Black victims yields,
$\hat{\theta}_{Y Z | X=1} = 0.68$ with $0.95$ CI $[ 0.28, 1.70]$ 
and $ \hat{\theta}_{Y Z | X=2}  = 0$ with $0.95$ CI $[0, 10.72]$.
To construct an exact test for $H_0$ the $217$ data sample points are ordered according to a statistic that quantifies their strength of evidence
in favor of Simpson's paradox,
and then the exact significance level  of the observed table is the sum of the probabilities of the data 
points with greater or equal test statistic value.
However, as Simpson's paradox involves effects having conflicting signs, determining strength of evidence in
favor of Simpson's paradox is difficult.
For example, does data point $(20, 0)$ with larger or equal conditional  associations 
( $\hat{\theta}_{Y Z | X=1} = 0.810$, $\hat{\theta}_{Y Z | X=2}  =  0$)  
and larger marginal ($\hat{\theta}_{Y Z}  =  1.34$) association 
offer more evidence in favor of Simpson's paradox than the observed data point?

\medskip
We propose two statistics for ordering the points in the data sample space. 
The first statistic is the posterior probability of the event corresponding to $H_1$
\[ {\cal P}_1 =  \{  ( \pi_{111} \cdots \pi_{2 2 2}) : \; \theta_{ Y  Z | X = 1} < 1, \; \theta_{ Y  Z | X = 2} < 1, \; 1 < \theta_{ Y  Z}  \ \}. \]
The second statistic is the ratio between the posterior probability of  ${\cal P}_1$ and the posterior probability of the event
\[ {\cal P}_0 (\epsilon)  =  \{  ( \pi_{111} \cdots \pi_{2 2 2}) : \;  | log(\theta_{ Y  Z | X = 1})| \le \epsilon  , \;   | log(\theta_{ Y  Z | X = 2})| \le \epsilon   \ \}, \]
with $\epsilon = 0.1$.
For our analysis we use a Dirichlet prior with concentration parameters $(0.5  \cdots  \; 0.5)$.
Thus for data point $( N_{1 1 1} \cdots \; N_{2 2 2} )$,
the posterior distribution of $( \pi_{1 1 1} \cdots \; \pi_{2 2 2} )$ is
Dirichlet with concentration parameters   $(N_{1 1 1} + 0.5   \cdots  \;  N_{2 2 2} + 0.5)$.
To compute the probability of  $ {\cal P}_1$ and  $ {\cal P}_0 (0.1) $ for a given data point,
we sample $( \pi_{1 1 1}, \cdots \pi_{2 2 2} )$ from the posterior probability and count the proportion
of samples that either events occurred.

Based on $2 \times 10^6$ samples from the posterior distribution,
data point  $( 20, 0)$ with $\Pr_{H_0} ( 20, 0 ) = 0.087$ 
has the largest  posterior probability of ${\cal P}_1$,
$0.085954$ ($s.e. < 0.0001$);
the observed table with $\Pr_{H_0} ( 19, 0 ) = 0.064$ has the second largest posterior probability,  $0.0797$ ($s.e. < 0.0001$);
Data point  $(  21, 0)$ with $\Pr_{H_0} ( 21, 0 ) = 0.101$ 
has the third largest  posterior probability, $0.0795$ ($s.e. < 0.0001$).
Thus for the first statistic, the significance level of the observed table is $0.151 = 0.087 + 0.064$.
To assess the posterior probability of $ {\cal P}_0 (\epsilon) $ we sampled $10^6$ realizations from the posterior distribution.
The posterior probability for the observed data point was $0.0054$.
Higher posterior probability was observed in $8$ data points, among them $(20, 0)$ and $(21,0)$.
In $121$ data points the ratio between the posterior probability of ${\cal P}_1$ and ${\cal P}_0 (0.1)$ was at least as high
as that of $(19,0)$, $14.8 = 0.0797 / 0.0054$.
The significance level of the observed table for the second statistic is $0.140$, 
the sum of the probabilities under the null for these $121$ data points.
\end{exa}

\begin{table}[ht]
\centering
\begin{tabular}{cc|cc}
 	Victim		& Defendant   		 & Death Penalty	& No Death Penalty \\  \hline
   White 			& White		 	& 19 			& 132 			\\ 
     		 		& Black 		 	& 11 			& 52 			\\  \hline
        Black		& White 		 	& 0				& 9				\\ 
   				&  Black 		 	& 6 				& 97 			\\ 
   \hline
\end{tabular}
  \caption{Death Penalty data} \label{table2}
  \end{table}


\bigskip
In Section 2 we present our general testing methodology and its conditional variant, 
phrase and prove their optimality property,
and explain the relation between our tests and Bayesian FDR controlling tests, Bayes factors
and likelihood ratio tests.
In Section 3 we demonstrate our methodology on a 4-by-4 contingency table,
present an exact tests for discovering positive dependence between two ordinal variables,
and perform a simulation that reveals that our methods may provide a slight power edge
even for testing composite null hypothesis that have a natural statistic.
We end the paper with a discussion.

\section{Mean most powerful tests}

We denote the parameter by ${\bs p} \in {\cal P}$, $\pi( {\bs p})$ is the prior distribution,
the data is ${\bs N} \in \Omega$, and the likelihood is $\Pr( {\bs n} | \ {\bs p})$.
The alternative hypothesis is $H_1: {\bs p} \in {\cal P}_1$, for ${\cal P}_1 \subseteq  {\cal P}$.
Following Benjamini and Hochberg (1995) that referred to rejecting the null hypothesis as making a statistical discovery,
${\cal P}_1$ is the discovery event and  we call ${\cal P}_0 \subseteq  {\cal P} - {\cal P}_1$ the non-discovery event.
The role of ${\cal P}_0$ is to determine the optimality property of the test, given in Definition \ref{dfn00}.
We explain how to set ${\cal P}_0$ in Remark \ref{rem00}.
The null hypothesis $H_0$ does not have to correspond to an explicit subset 
of ${\cal P}_0$,
all we will need is that the null hypothesis specifies a null distribution 
$\Pr_{H_0} ({\bs N} = {\bs n}  )$ on $\Omega$.
Tests are mappings ${\cal T} : \ \Omega \rightarrow \{ 0, 1 \}$.
For $S \subseteq \Omega$, let 
${\cal T} (S) : = I ( {\bs n} \in S)$,
where ${\cal T} (S) = 1$ corresponds to declaring that  ${\bs p} \in {\cal P}_1$.
Thus the  significance level of ${\cal T} (S)$ is
$Pr_{H_0} ( {\bs N} \in S)$.

\bigskip
Our tests are Bayes rules for discriminating between ${\cal P}_0$ and ${\cal P}_1$ 
that minimize the average risk for the following loss function:
\begin{equation} \label{loss-func}
L ( S  ; \lambda_1, \lambda_2 ) = 
\lambda_1 \cdot  I ( {\bs N} \in S , \ {\bs P} \in {\cal P}_0 ) + 
\lambda_2 \cdot  I ( {\bs N} \notin S , \ {\bs P} \in {\cal P}_1 ).
\end{equation}
As the marginal distribution of ${\bs N}$ is
 \[
\Pr( {\bs N} = {\bs n} ) = \int_{\bs p} \pi (\bs p) \cdot  \Pr ( {\bs N} = {\bs n} | \ {\bs p} ) \  d {\bs p},
 \]
 and the conditional distribution of ${\bs p}$ given ${\bs N} = {\bs n}$ is
 \[
\pi ( {\bs p} | \  {\bs n}) =   \Pr ( {\bs N} = {\bs n} | \ {\bs p} )  \cdot  \pi (\bs p) /  \Pr( {\bs N} = {\bs n} ),
 \]
the average risk can be expressed
 \begin{eqnarray} 
\lefteqn{  \sum_{{\bs n} \in \Omega} \Pr(  {\bs n} ) \cdot  \int_{{\bs p}}  \pi ({\bs p} | \ {\bs n}  ) 
\cdot [ \lambda_1 \cdot  I ( {\bs n} \in S , \ {\bs P} \in {\cal P}_0 ) + 
\lambda_2 \cdot  I ( {\bs n} \notin S , \ {\bs P} \in {\cal P}_1 ) ]  \ d {\bs p}} \nonumber \\
&   = &   
\sum_{{\bs n} \in S}  \Pr(  {\bs n} ) \cdot  \lambda_1 \cdot \Pr( {\bs P} \in {\cal P}_0 | \ {\bs n}  ) + \
\sum_{{\bs n} \notin S}  \Pr(  {\bs n} ) \cdot  \lambda_2 \cdot \Pr( {\bs P} \in {\cal P}_1 | \ {\bs n}  ). \label{ave-risk}
 \end{eqnarray}
Thus for $\delta = \lambda_1 / \lambda_2$, $S$ that   minimizes the average risk in (\ref{ave-risk}) is
\begin{equation} \label{BFdef}
S^{Bayes}  ( \delta) = \{ {\bs n} : \  \delta  \le     
\frac{\Pr( {\bs P} \in {\cal P}_1 | \ {\bs n}  ) }{\Pr( {\bs P} \in {\cal P}_0 | \ {\bs n} )} \}.
\end{equation}
 
 \noindent Similarly, the Bayes rule can be specified according to its significance level.
 For $\alpha \in [0,1]$, let $S^{Bayes} (\alpha) := S^{Bayes} (\delta_{\alpha})$
 for 
 \[\delta_{\alpha} = \min \{ \delta : \;   Pr_{H_0} ( {\bs N} \in S^{Bayes} ( \delta)) \le \alpha  \  \}. \]

\begin{dfn} \label{dfn00}
\begin{enumerate}
\item The {\em mean significance level} of ${\cal T} (S)$ is
$Pr( {\bs N} \in S | \ {\bs p} \in {\cal P}_0)$.
\item The {\em mean power} of ${\cal T} (S)$ is
$Pr( {\bs N} \in S | \ {\bs p} \in {\cal P}_1)$.
\item ${\cal T} (S)$ is a {\em mean most powerful}  test  if 
all tests with less or equal mean significance level have less or equal mean power.
\end{enumerate}
\end{dfn}

\medskip
\begin{prop}  \label{prop1}
$\forall \delta$,  ${\cal T} (S^{Bayes} (\delta) )$ is  a mean most powerful test.
\end{prop}

\begin{pf}
Let ${\cal T} ( \tilde{S})$ be a test with  less or equal  mean significance than  ${\cal T} (S^{Bayes})$,
\begin{equation} \label{eq3b}
\Pr( {\bs N} \in \tilde{S} | \  {\bs P} \in {\cal P}_0) 
\le Pr( {\bs N} \in S^{Bayes}  | \ {\bs P} \in {\cal P}_0).
  \end{equation}

\medskip \noindent We begin by expressing
\begin{equation} \label{eq3c}
\Pr( {\bs N} \in \tilde{S} | \  {\bs p} \in {\cal P}_0) = 
 \sum_{ {\bs n} \in \tilde{S}} 
 \Pr(    {\cal P}_0 | \   {\bs n} ) \cdot  \Pr(  {\bs n} ) /  \Pr(   {\cal P}_0),
\end{equation}
and expressing 
\begin{equation} \label{eq3}
\Pr( {\bs N} \in S^{Bayes} | \  {\bs p} \in {\cal P}_0) =
 \sum_{  {\bs n} \in S^{Bayes}}   \Pr(   {\cal P}_0 | \  {\bs n}  ) \cdot  \Pr(  {\bs n} )    /  \Pr(   {\cal P}_0). 
  \end{equation}
 Subtracting the summands in $S^{Bayes} \cap \tilde{S}$ from the sums in (\ref{eq3c}) and (\ref{eq3})
 and multiplying by $\Pr( {\cal P}_0)$,
Inequality (\ref{eq3b}) implies that  
\begin{equation} \label{eq3e}
\sum_{ {\bs n} \in \tilde{S} - (S^{Bayes} \cap \tilde{S})} 
 \Pr(   {\cal P}_0 | \ {\bs n} ) \cdot  \Pr( {\bs n} ) 
 \le 
 \sum_{ {\bs n} \in S^{Bayes} -(S^{Bayes} \cap \tilde{S})} 
 \Pr(    {\cal P}_0 | \   {\bs n} ) \cdot  \Pr(  {\bs n} ). \label{ineq4a}
  \end{equation}
 According to the construction of $S^{Bayes}$,  $\forall {\bs n}_1 \in \tilde{S} -  (S^{Bayes} \cap  \tilde{S})$
and $\forall {\bs n}_2 \in S^{Bayes} -  (S^{Bayes} \cap  \tilde{S})$ 
\begin{equation} \label{ineq4e}
 \Pr(   {\cal P}_1 | \  {\bs n}_1 ) /  \Pr(   {\cal P}_0 | \  {\bs n}_1 ) \le
 \Pr(   {\cal P}_1 | \  {\bs n}_2 ) / \Pr(   {\cal P}_0 | \  {\bs n}_2 ).
   \end{equation}
 Next, we express
 \begin{eqnarray}
\lefteqn{\Pr( {\bs N} \in \tilde{S} | \  {\bs p} \in {\cal P}_1) =  \sum_{ {\bs n} \in S^{Bayes} \cap \tilde{S}} 
   \Pr(   {\cal P}_1 | \  {\bs n}  ) \cdot  \Pr(  {\bs n} )    /  \Pr(   {\cal P}_1)} \label{ineq8d}  \\
 & +  & 
 \sum_{ {\bs n} \in \tilde{S} -(S^{Bayes} \cap \tilde{S})} 
 (  \Pr(  {\cal P}_0 | \   {\bs n} ) \cdot    \frac{\Pr(  {\cal P}_1 | \   {\bs n} )}{\Pr(  {\cal P}_0 | \   {\bs n} )} )  \cdot \frac{\Pr( {\bs n} ) }{ \Pr(   {\cal P}_1)}  . \label{ineq8a}
 \end{eqnarray}
 and 
  \begin{eqnarray}
\lefteqn{\Pr( {\bs N} \in \tilde{S} | \  {\bs p} \in {\cal P}_1) =  \sum_{ {\bs n} \in S^{Bayes} \cap \tilde{S}} 
   \Pr(   {\cal P}_1 | \  {\bs n}  ) \cdot  \Pr(  {\bs n} )    /  \Pr(   {\cal P}_1)}\label{ineq8c}  \\
 & +  & 
 \sum_{ {\bs n} \in S^{Bayes} -(S^{Bayes} \cap \tilde{S})} 
 (  \Pr(  {\cal P}_0 | \   {\bs n} ) \cdot    \frac{\Pr(  {\cal P}_1 | \   {\bs n} )}{\Pr(  {\cal P}_0 | \   {\bs n} )} )  \cdot \frac{\Pr( {\bs n} ) }{ \Pr(   {\cal P}_1)}.   \label{ineq8b}
 \end{eqnarray}
 Note that Expression (\ref{ineq8a}) is the left hand side of (\ref{ineq4a})
 and Expression (\ref{ineq8b}) is the right hand side of (\ref{ineq4a}), divided by $\Pr({\cal P}_1)$
 and multiplied by a factor, that according to (\ref{ineq4e}), is larger in each summand of (\ref{ineq8b})
 than in all of the summands of (\ref{ineq8a}). 
 Therefore the sum in  (\ref{ineq8b}) is larger than the sum in  (\ref{ineq8a}),
 and as the sums in the right hand side of  (\ref{ineq8d}) and  (\ref{ineq8c}) are the same,
 \[ \Pr( {\bs N} \in \tilde{S} | \  {\bs p} \in {\cal P}_1) \le \Pr( {\bs N} \in S^{Bayes} | \  {\bs p} \in {\cal P}_1).
 \]
 \end{pf}
 
 \begin{rem} \label{rem00}
Determining ${\cal P}_1$, ${\cal P}_0$, and $\pi( {\bs p})$, produces a family of mean most powerful tests.
Per construction, ${\cal T}(S^{Bayes} (\alpha))$ has significance level $\alpha$ and has more mean power than 
all mean most powerful tests with significance level  $< \alpha$. According to Proposition \ref{prop1},
${\cal T}(S^{Bayes} (\alpha))$ also has more mean power than all tests will smaller or equal mean significance level.
Note that in the examples in the paper we only compute the p-value for the observed data, applying
${\cal T}(S^{Bayes} (\alpha))$ further entails rejecting $H_0$ if the p-value is $\le \alpha$.

Ideally,  the prior distribution captures the knowledge regarding the parameters
that is available prior to the study.
In the examples in the paper we used conjugate non-informative priors that provide easy test statistic computation
and yield general optimal tests for each alternative null hypothesis.
While the choice of ${\cal P}_1$ is usually dictated by the application,
 ${\cal P}_0$ can just be a subset of ${\cal P} - {\cal P}_1$. 
We suggest either setting ${\cal P}_0$ to be 
a ``small'' set containing ${\bs p}_0$, the parameter value under the null  (we denoted this set by ${\cal P}_0 (\epsilon)$ in Example \ref{exa11}),
or setting ${\cal P}_0 = {\cal P} - {\cal P}_1$.
If ${\cal P}_0 = \{ {\bs p}_0 \}$,
then the mean significance level would equal the significance level, 
thus ${\cal T} (S^{Bayes} (\alpha))$ would have more mean power then all tests with significance level $\le \alpha$.
As our choice of priors assigns zero probability to $\{ {\bs p}_0 \}$, we resort to setting ${\cal P}_0  = {\cal P}_0 (\epsilon)$
with small $\epsilon$ that produces a very similar family of mean most powerful tests.
But note that using too small  $\epsilon$ will make it very difficult to numerically assess $ \Pr( {\cal P}_0 (\epsilon) | \ {\bs n} )$.
The other option is setting ${\cal P}_0 = {\cal P} - {\cal P}_1$, that yields
\[    
\frac{\Pr( {\bs P} \in {\cal P}_1 | \ {\bs n}  ) }{\Pr( {\bs P} \in {\cal P}_0 | \ {\bs n} )}  = 
\frac{\Pr( {\bs P} \in {\cal P}_1 | \ {\bs n}  ) }{1 - \Pr( {\bs P} \in {\cal P}_1 | \ {\bs n} )}. 
\]
This means that sorting the data points according to $Pr(  {\cal P}_1 | \ {\bs n} )$
is equivalent to sorting the data points according to $Pr(  {\cal P}_1 | \ {\bs n}  ) /  \Pr( {\cal P}_0 | \ {\bs n} )$.
In this case the optimality property may be less appealing but it has the great technical advantage
that to construct our test, for each data point, we only need to assess the posterior probability of ${\cal P}_1$. 
  \end{rem}

\subsection{Conditional mean most powerful tests}

In this section we present mean most powerful tests for the conditional analysis of contingency tables, 
in which the sample space is partitioned according to the row and column sums 
and a separate level $\alpha$ test is conducted in each partition. 

\medskip
Let $a$ be the statistic that partitions the sample space $\Omega = \cup_{a \in {\cal A}} \Omega_{a}$,
for ${\cal A} = \{ a ( N) : \; N \in \Omega \}$ the set of statistic values.
\begin{dfn}
A conditional level $\alpha$ test 
is ${\cal T} ( {\cal S}_{\cal A} (\alpha) )$ such that $\forall a \in {\cal A}$,
$\Pr_{H_0} ( {\bs N} \in {\cal S}_{\cal A} (\alpha) | {\bs N} \in   \Omega_{a} ) \le \alpha$.
\end{dfn}
\noindent 
To construct ${\cal S}^{Bayes}_{\cal A} (\alpha)$, the rejection region of the conditional mean most powerful test,
we repeat the following for each $a \in {\cal A}$ :
 sort  the data points ${\bs N} \in \Omega_a$ according to $\Pr ( {\bs P} \in {\cal P}_1 | {\bs N}) / \Pr ( {\bs P} \in {\cal P}_0 | {\bs N})$
 and then following that order, 
 as long as $\Pr_{H_0} ( {\bs N} \in {\cal S}^{Bayes}_{\cal A} (\alpha) | \ {\bs N} \in   \Omega_{a} ) \le \alpha$,
 sequentially add data points into ${\cal S}^{Bayes}_{\cal A} (\alpha)$.
 
\begin{rem}
Per construction, ${\cal T} (S^{Bayes}_{\cal A}(\alpha))$ is a conditional level $\alpha$ test
and for all $a$, ${\cal T} (S^{Bayes}_{\cal A}(\alpha) \cap \Omega_a )$  is a mean most powerful test on $\Omega_a$.
Conditional level $\alpha$ tests are also  level $\alpha$ tests:
 \begin{eqnarray*}
\lefteqn{\Pr_{H_0} ( {\bs N} \in {\cal S}_{\cal A} (\alpha)  )  =  \sum_{a \in {\cal A} }\Pr_{H_0} (  {\bs N} \in {\cal S}_{\cal A} (\alpha),   {\bs N}\in   \Omega_{a} )}  \\
 & = & \sum_{a \in {\cal A} } \Pr_{H_0} (  {\bs N} \in {\cal S}_{\cal A} (\alpha) |  {\bs N} \in   \Omega_{a} ) \cdot \Pr_{H_0} (  {\bs N} \in   \Omega_{a} )  
  \le   \sum_{a \in {\cal A} } \alpha  \cdot \Pr_{H_0} ( \ N \in   \Omega_{a} )  \; = \; \alpha.
\end{eqnarray*}
When $a$ assumes a single value  then ${\cal S}^{Bayes}_{\cal A} (\alpha) =  {\cal S}^{Bayes} (\alpha)$.
But in general,  ${\cal T} (S^{Bayes}_{\cal A}(\alpha))$ is not a mean most powerful test and there may even be other 
conditional level $\alpha$ test with smaller mean significance level and larger mean power.
However, if ${\cal P}_0 = \{ {\bs p}_0 \}$ and $\Pr_{H_0} ( {\bs N} \in {\cal S}^{Bayes}_{\cal A} (\alpha) | \ {\bs N} \in  \Omega_{a} )  = \alpha$ for all $a$,
then as ${\cal T} (S^{Bayes}_{\cal A}(\alpha) \cap \Omega_a )$  is a mean most powerful test on $\Omega_a$
and the mean significance level identifies with the significance level, 
any other conditional level $\alpha$ test,  ${\cal T} (S_{\cal A}(\alpha))$,  
would have smaller mean significance level than ${\cal T} (S^{Bayes}_{\cal A}(\alpha))$ on $\Omega_a$ and thus  
 it would also have smaller mean power on $\Omega_a$.
Summing over all $\Omega_a$, ${\cal T} (S_{\cal A}(\alpha))$ would have smaller mean power than 
${\cal T} (S^{Bayes}_{\cal A}(\alpha))$.
 \end{rem}

\subsection{Relation between our tests and Bayesian FDR controlling tests,
Bayes factors, and likelihood ratio tests}

$\Pr( {\bs P} \in {\cal P}_1 | {\bs n} )$ is equal to one minus the local FDR (Efron et al., 2001).
Thus setting ${\cal P}_0 = {\cal P} - {\cal P}_1$ we follow Storey (2007), who suggested constructing optimal tests
 in which the local FDR is used for determining  the order in which the data points are included into the rejection region.
 However, unlike the Bayesian FDR approach, in which $\pi ({\bs p})$ is the marginal parameter
 distribution in the population of parameters that is under study, and thus the Bayesian FDR 
 can be used to determine  the cutoff point of the rejection region  (Heller and Yekutieli, 2012).
 In our tests the cutoff point is determined by the test's significance level.
 
\medskip
Expressing the statistic in (\ref{BFdef})
\begin{equation}
\frac{\Pr( {\bs P} \in {\cal P}_1 | \ {\bs N} = {\bs n}  ) }{\Pr( {\bs P} \in {\cal P}_0 | \ {\bs N} = {\bs n} )}  =
\frac{ \frac{\Pr( {\bs N} = {\bs n}  | \ {\bs P} \in {\cal P}_1 ) \cdot \Pr(  {\bs P} \in {\cal P}_1 )}
{\Pr( {\bs N} = {\bs n}) } }
{ \frac{\Pr( {\bs N} = {\bs n}  | \ {\bs P} \in {\cal P}_0 ) \cdot \Pr( {\bs P} \in {\cal P}_0 )}
{\Pr( {\bs N} = {\bs n}) }}  \propto
\frac{\Pr( {\bs N} = {\bs n}  | \ {\bs P} \in {\cal P}_1 )}
{\Pr( {\bs N} = {\bs n}  | \ {\bs P} \in {\cal P}_0 )},  \label{eq445}
\end{equation}
reveals that we actually order the data points according to the Bayes factor between  ``model'' ${\cal P}_1$ and ``model'' ${\cal P}_0$.
However, note that in our tests the cutoff point of the rejection region 
is not a nominal Bayes factor value (cf. Kass and Raftery, 1995).

\medskip
Our tests are also closely related to likelihood ratio tests.
For simple hypotheses, $H_0: {\bs p} = {\bs p}_0$ for ${\bs p}_0 \in {\cal P}_0$ vs. $H_1: {\bs p} = {\bs p}_1$ for ${\bs p}_1 \in {\cal P}_1$, 
our test reduces to the likelihood ratio
test if ${\cal P}_0 = \{ {\bs p}_0 \}$ and ${\cal P}_1 = \{ {\bs p}_1 \}$,
or if the prior distribution assigns all its probability to the two hypotheses:
$\pi({\bs p}_0) = \pi_0$ and  $\pi({\bs p}_1) = 1 - \pi_0$, for $0 < \pi_0  < 1$.
The likelihood ratio statistic (Casella and Berger,  2001) for testing the composite hypotheses 
$H_0: {\bs p} \in {\cal P}_{null}$ 
vs.  $H_1 : {\bs p} \notin {\cal P}_{null}$ is
\[
\Lambda ( {\bs n} )  = \frac{ \sup_{ {\bs p} \in {\cal P}_{null}} \Pr ( {\bs N} = {\bs n} | {\bs p} ) }
{ \sup_{ {\bs p} \in {\cal P}} \Pr ( {\bs N} = {\bs n} | {\bs p} ) }.
\]
For ${\cal P}_1  = {\cal P} - {\cal P}_{null}$, setting ${\cal P}_0 =  {\cal P} -  {\cal P}_1$ yields
${\cal P}_0 = {\cal P}_{null}$ and thus $\Lambda ( {\bs n} )$ orders the data points similarly to one minus our statistic,
except that  in our statistic we consider the average rather than the supremum of the likelihood, 
which according to our theoretical results yields tests with more power with respect to the prior distribution. 
However for ${\cal P}_1  \subset {\cal P} - {\cal P}_{null}$ and setting ${\cal P}_0 =  {\cal P} -  {\cal P}_1$,
our statistic, that orders the data points according to ${\cal P}_1$,
yields considerably more powerful  tests than $\Lambda ( {\bs n} )$, 
that orders the data points according to the null hypothesis, 
especially for the case that ${\cal P}_1$ is a ``small'' subset of ${\cal P} - {\cal P}_{null}$.
We illustrate this in the following example and it occurs in the two contingency table examples,
where our tests yield considerably smaller p-values than the $X^2$  statistic, 
which is the likelihood ratio statistic for testing independence for cross-tabulated data.

\begin{exa} \label{exa44}
The parameter is ${\bs \mu} =  (\mu_1 \cdots \mu_K)$.
The data is   ${\bs Y} = (Y_1 \cdots Y_K)$ with $Y_k \sim N(\mu_k, 1)$.
The null hypothesis is $H_0: {\mu} =  0$ and ${\cal P}_1 = \{  {\bs \mu} : \ 3 \le \mu_1 \}$. 
In the likelihood ratio test for $H_0: {\mu} =  0$ vs. $H_1: {\mu} \ne 0$,  
the data points are ordered according to their $l_2$ norm.
Setting ${\cal P}_0 = {\cal P} - {\cal P}_1$  and using a flat prior for ${\bs \mu}$,
our test sorts the data points are ordered according to $Y_1$. 
For $K = 100$ and ${\bs \mu} = (3.2, 0 \cdots 0)$,
as $124.34$ is the $0.95$ quantile of the $100$ degree of freedom  $\chi^2$ distribution,
the rejection region for the $\alpha = 0.05$ likelihood ratio test is $\{ {\bs y} : \ 124.34 \le \| {\bs y} \|^2 \}$ 
 and the power of this test is $0.179$,
while for our $\alpha = 0.05$ test the rejection region is $S^{Bayes} ( 0.05)  = \{ {\bs y} : \ 1.64 \le y_1  \}$  and its power is $0.940$.
\end{exa}

\section{Job Satisfaction Example}

The data in Table 2 was also taken from Agresti (2002, Table 2.8).
A sample of $96$  black males  were classified by Income
(``$<1500$'', ``$15000-25000$'', ``$25000-40000$'', ``$>40000$'')
and job satisfaction
 (``Very Dissatisfied'', ``Little Dissatisfied'', ``Moderately  Satisfied'', ``Very Satisfied'').
 For $i = 1 \cdots 4$ and $j = 1 \cdots 4$,
$\pi_{i j}$ is the probability that a respondent has income level $i$ and job satisfaction level $j$.
We assume that the number of respondents ${\bs N} = (N_{1 1} \cdots \;  N_{4 4})$ is $multinom (\pi_{1 1} \cdots \; \pi_{4 4})$. 
$n_{i j}$ is the observed number of respondents recorded in Table 2.
The null hypothesis is $H_0: \pi_{i j}  = \pi_{i +}  \pi_{+ j}$,
for $\pi_{i +} = \pi_{i 1} + \cdots + \pi_{i 4}$ and $\pi_{+ j} = \pi_{1 j} + \cdots + \pi_{4 j}$.
A pair of respondents is concordant if they have different income and job satisfaction and the 
respondent with higher income has higher job satisfaction.
The probability that a pair of respondents is concordant is 
\begin{equation} \label{eqn41}
\Pi_C = 2  \sum_i \sum_j \pi_{i j}  ( \sum_{i < h} \sum_{j < k} \pi_{h k} ).
\end{equation}
A pair of respondents is discordant if they have different income and job satisfaction and the 
respondent with higher income has lower job satisfaction.
The probability that a pair of respondents is discordant is 
\begin{equation} \label{eqn42}
\Pi_D = 2   \sum_i \sum_j \pi_{i j}  ( \sum_{i < h} \sum_{k < j} \pi_{h k} ).
\end{equation}
The degree of concordance is measured by Kendall's gamma rank correlation coefficient, 
$\gamma = (\Pi_C  - \Pi_D ) / ( \Pi_C  + \Pi_D )$.
Which is the difference between the conditional probability of concordance and discordance
given that the pair of respondents have different income and different job satisfaction.

\medskip 
We first test $H_0$ with tests implemented in R, whose
significance levels are based on parametric approximations of the test statistics'  distribution under the null hypothesis.
Pearson's Chi-squared test ({\em chisq.test} function) 
yielded $X^2 = 5.97$ with $9$ degrees of freedom and p-value $0.743$.
Kendall's rank correlation coefficient ({\em cor.test} function), corresponding to
alternative hypothesis of concordance between of income and job satisfaction, was
 $\tau = 0.152$ with p-value $0.043$.
 Spearman's rank correlation coefficient ({\em cor.test} function), corresponding to alternative hypothesis 
 of positive rank correlation, was $\rho = 0.177$ with p-value $0.042$.

To construct the exact tests we condition on $n_{i +}$ and $n_{+ j}$,
the row and column sums of Table 2.
There are $90,208,550$ possible 4-by-4 tables with the same row and columns sums as Table 2.
Under the null hypothesis, the distribution of these tables is multivariate hypergeometric.

The first exact test is based on Kendall's gamma estimator,
$\hat{\gamma} = (\hat{\Pi}_C  - \hat{\Pi}_D ) / ( \hat{\Pi}_C  + \hat{\Pi}_D )$,
for $\hat{\Pi}_C$  and  $\hat{\Pi}_D$ computed by replacing 
$\pi_{i j}$ with  $\hat{\pi}_{i j} = N_{i j} / 96$ in (\ref{eqn41}) and (\ref{eqn42}).
The observed value is $\hat{\gamma} = 0.221$.
Greater or equal $\hat{\gamma}$ values were computed for $21,101,151$ tables.
The sum of the probabilities under $H_0$ of these tables was $0.0415$.

Our second statistic is the posterior probability of the concordance event, 
 ${\cal P}^{Cncrd}_1 = \{ ( \pi_{1 1} \cdots \pi_{4 4})  : \ 0 \le \gamma \}$.
We use a Dirichlet prior  distribution with concentration parameters $(0.5  \cdots  \; 0.5)$
for $( \pi_{1 1} \cdots \; \pi_{4 4} )$, for which the posterior probability is a dirichlet distribution with  concentration parameters 
$(N_{1 1} + 0.5   \cdots  \;  N_{4 4} + 0.5)$. 
To compute the probability of the concordance event for a given table,
we sample $( \pi_{1 1}, \cdots \pi_{4 4} )$ from the posterior probability and record the proportion
of times the concordance event occurs.
The probability of concordance for $N_{i j} = n_{i j}$, based on a
 sample of $10^7$ draws from the posterior, was $0.9564$ ($s.e. < 0.0001$).
 Computing this statistic for all 4-by-4 tables is too time consuming. 
 Thus to assess the significance level  for this statistic,
  we generated a sample of $50,000$ 4-by-4 contingency tables
 from  the multivariate hypergeometric null distribution, and for each contingency table we sampled $10,000$ 
 $( \pi_{1 1}, \cdots \pi_{4 4} )$ from the posterior probability and recorded the proportion
of times the concordance event occurred. The estimated significance level was $0.036$ 
($s.e. < 0.001$), the proportion of contingency tables with estimated proportion of concordance $\ge 0.9564$.

Our statistic for the third exact test is the posterior probability that income and job satisfaction
are positively dependent.
This is a stronger property than concordance that corresponds to the event
\begin{equation} \label{eqn43}
{\cal P}^{Pos}_1 = \{  ( \pi_{1 1}, \cdots,  \pi_{4 4} ) : \; \Pr ( \pi_{ j | i} \le t ) \ge  \Pr ( \pi_{ j | i + 1} \le t ) 
\;  \; \forall t, \forall j, \forall i  \},
 \end{equation}
 for $\pi_{ j | i} = \pi_{i j} / \pi_{i +}$. 
 Based on a sample of $10^7$ draws,
the posterior  probability of positive dependence for the observed table is $0.0118$  ($s.e. < 0.0001$).
And again, to assess the significance level for this statistic we  sampled $50,000$ 4-by-4 contingency tables
 from  the multivariate hypergeometric null distribution
 and for each contingency table we sampled $10,000$  $( \pi_{1 1}, \cdots \pi_{4 4} )$ from the posterior probability.
The estimated significance level was $0.0093$  ($s.e. < 0.001$), 
which is the proportion of contingency tables with posterior probability of positive dependence 
$\ge 0.0118$.

Note that for the two Bayesian statistics we set ${\cal P}_0 = {\cal P} - {\cal P}_1$. 
For ${\cal P}_1 = {\cal P}^{Pos}_1$, 
we had also experimented with setting ${\cal P}_0$ to be  a small subset containing the null, ${\cal P}_0 (\epsilon)$.
However, with $\epsilon$ large enough to be able to estimate the posterior probability of ${\cal P}_0 (\epsilon)$
in comparable run time the p-value increased from less than $1\%$ to more than $10\%$,
suggesting that for this data setting ${\cal P}_0 = {\cal P}_0 (\epsilon)$ is not a feasible option.
\bigskip

\begin{table}[ht]
\begin{center}
\begin{tabular}{ccccc}
& \multicolumn{4}{c}{Job Satisfaction}  \\
  \cline{2-5}
 Income & Very & Little  & Moderately & Very \\ 
(Dollars) & Dissatisfied & Dissatisfied & Satisfied &  Satisfied\\ 
  \hline
$<$15000 & 1  & 3  & 10  & 6  \\ 
  15000-25000 & 2  & 3  & 10  & 7  \\ 
  25000-40000 & 1  & 6  & 14  & 12  \\ 
  $>$40000 & 0  & 1  & 9  & 11  \\ 
   \hline
 \end{tabular}
\end{center}
  \caption{Job Satisfaction data} \label{table1}
\end{table}

\subsection{Job Satisfaction Simulation}
The simulation compares the power of the conditional exact test whose  test statistic is  $\hat{\gamma}$
with the conditional exact test whose test statistic is $\Pr( 0 \le \gamma |  \ N_{1 1}  \cdots \; N_{4 4} )$, on 
\begin{equation} \label{sampspc}
\Omega_a = \{ ( N_{1 1}  \cdots \; N_{4 4}):  \ N_{1 +} = n_{1 +}, N_{2 +} = n_{2 +},  \cdots , N_{+ 4} = n_{+ 4} \}
\end{equation}
for which the null distribution of ${\bs N}$ is the multivariate hypergeometric considered in the previous section.
The alternative distribution is that ${\bs N}$ is    
$multinomial$ $( \hat{\pi}_{1 1}  \cdots \;  \hat{\pi}_{4 4} )$, with  $\hat{\pi}_{i j} = n_{i j} / 96$,
truncated to $\Omega_a$ in (\ref{sampspc}).

We use importance sampling to generate ${\bs N}$ from the alternative distribution.
We sample $10^6$ proposal realizations  of ${\bs N}$ from the multivariate hypergeometric
null distribution; for each proposal realization we compute a sampling weight that is the probability 
of observing this realization under the alternative multinomial distribution divided by the probability of observing this
realization under the multivariate hypergeometric null distribution;
and use weighted with-replacement sampling of the $10^6$ proposal values to
generate a sample of $10^5$ realizations  from the alternative distribution.
We then compute the two test statistic values for each of the $10^5$ realizations.
Lastly, to assess the significance level of  each realization for the two test statistics,
we generate another sample of $10^5$ realizations of $( N_{1 1}  \cdots \; N_{4 4} )$ from the null distribution,
and compute the two test statistic value for each null realization.
The p-values assigned to each alternative distribution realization is the 
proportion of null realization for which the statistic values were larger than the 
realization's test statistic values.

Recall that for the Table 2 data, the p-value for the exact test based 
on the $\hat{\gamma}$ statistic was $0.0415$ and the p-value for the exact test for the probability of concordance
statistic was $0.036$.
In our simulation, for the $\hat{\gamma}$ statistic computed for the $10^5$ alternative distribution realizations,
 the mean p-value was $0.0988$ and the median p-value was $0.0399$,
 $0.679$ ($s.e. < 0.005$) of the p-values were smaller then $0.10$ 
 and $0.537$ ($s.e. < 0.005$) of the p-values were smaller than $0.05$.
While for the p-values computed based on the probability of concordance statistics,
 the mean p-value was $0.0947$ and the median p-value was $0.0370$,
 $0.701$ ($s.e. < 0.005$) of the p-values were smaller then $0.10$ 
 and $0.550$ ($s.e. < 0.005$) of the p-values were smaller than $0.05$.

\section{Discussion}

As we will usually need to assess our test statistic values and their significance levels
numerically by simulation from the null hypothesis, followed by simulation from the 
parameter posterior distribution, our tests can be computationally intensive.
We therefore suggest using our tests in ``difficult'' cases where the parameter space
is high dimensional and we know how to express the alternative hypothesis as a subset
of the parameter space, however it is not clear how to construct a test statistic for this hypothesis.
We also suggest using our methods in cases where there is prior information
on the parameter or for very high dimensional and very sparse tables in which the 
asymptotic results for the test statistic distribution fail and the usual statistics may be severely under powered. 

We presented methodology for the analysis of contingency tables in which use of exact tests is well established.
However note that our approach can also be used  to construct optimal tests
for other problems in which samples under the null hypothesis can be generated by permutations or bootstrapping.


\begin{thebibliography}{}

\bibitem{KR}  Agresti A.  (2002) ``Categorical Data Analysis`` John Wiley \& Sons, 2002.


\bibitem{SS} Benjamini Y., Hochberg Y. (1995) `Controlling the false discovery rate: a practical and powerful approach to multiple testing'' 
{\em J. R. Statist. Soc. B}, {\bf 57}, 289-300.

\bibitem{CB} Casella G., Berger R. L. (1990)  ``Statistical inference'' Belmont, CA: Duxbury Press.


\bibitem{ET} Efron B., Tibshirani R., Storey J. D. and Tusher V. (2001) 
``Empirical Bayes analysis of a microarray experiment''.
{\em J. Am. Statist. Ass.}, {\bf 96}, 1151Ð1160.

\bibitem{HY} 	Heller R.,  Yekutieli D.,   ``Replicability analysis for genome-wide association studies''
{\em arXiv:1209.2829}

\bibitem{KR}  Kass R. E., Raftery A. E. (1995) ``Bayes Factors'' {\em Journal of the American Statistical Association},  {\bf  90},  773-795.

\bibitem{SS} Storey, J D. (2007) ``The optimal discovery procedure: a new approach to simultaneous significance testing'' 
{\em J. R. Statist. Soc. B}, {\bf 69}, 347Ð368.

\end{thebibliography}
\end{document}